\documentclass[reprint, secnumarabic,amssymb, nobibnotes, superscriptaddress, aps, prl]{revtex4-1}
\usepackage{graphicx}
\usepackage[colorlinks=true,urlcolor=blue]{hyperref}
\usepackage{hyperref}
\usepackage{xcolor}
\hypersetup{
    colorlinks,
    linkcolor={red!50!black},
    citecolor={blue!50!black},
    urlcolor={blue!80!black}
}

\begin{document}

\title{Surface collective modes in the topological insulators Bi$_2$Se$_3$ and Bi$_{0.5}$Sb$_{1.5}$Te$_{3-x}$Se$_{x}$}%

\author{A. Kogar}
\author{S. Vig} 
\author{A. Thaler} 
\author{M.H. Wong}
\author{Y. Xiao}
\author{D. Reig-i-Plessis}

\affiliation{\mbox{Department of Physics and Seitz Materials Research Laboratory, University of Illinois, Urbana, IL, 61801, USA}}
\author{G.Y. Cho}
\affiliation{\mbox{Department of Physics, Korea Advanced Institute of Science and Technology, Daejeon 305-701, Korea}}
\author{T. Valla}
\author{Z. Pan} 
\author{J. Schneeloch} 
\author{R. Zhong} 
\author{G.D. Gu}
\affiliation{\mbox{Condensed Matter Physics and Materials Science Department, Brookhaven National Laboratory, Upton, New York 11973, USA}}
\author{T.L. Hughes}
\author{G.J. MacDougall}
\author{T.-C. Chiang}
\author{P. Abbamonte}
\email{abbamonte@mrl.illinois.edu}
\affiliation{\mbox{Department of Physics and Seitz Materials Research Laboratory, University of Illinois, Urbana, IL, 61801, USA}}
\date{\today}

\begin{abstract}
We used low-energy, momentum-resolved inelastic electron scattering to study surface collective modes of the three-dimensional topological insulators 
Bi$_2$Se$_3$ and Bi$_{0.5}$Sb$_{1.5}$Te$_{3-x}$Se$_{x}$. Our goal was to identify the ``spin plasmon" predicted by Raghu and co-workers [S. Raghu, et al., Phys. Rev. Lett. {\bf 104}, 116401 (2010)]. Instead, we found that the primary collective mode is a surface plasmon arising from the bulk, free carrers in these materials. This excitation dominates the spectral weight in the bosonic function of the surface, $\chi ''(\textbf{q},\omega)$, at THz energy scales, and is the most likely origin of a quasiparticle dispersion kink observed in previous photoemission experiments. Our study suggests that the spin plasmon may mix with this other surface mode, calling for a more nuanced understanding of optical experiments in which the spin plasmon is reported to play a role.  
\end{abstract}

\maketitle

The defining characteristics of the three-dimensional topological insulators are a bulk gap and the presence of surface states that cannot be gapped by any time-reversal symmetric type of disorder\cite{KaneMele, BHZ, zhang2009topological, roy2009, mooreBalents}. At its $\Gamma$ point, the prototypical Bi$_2$Se$_3$ system exhibits a single surface Dirac cone characterized by locking between the quasiparticle spin and momentum\cite{hsieh2008ARPES, ChenBi2Te3, Hsieh2009ARPES, xia2009ARPES, roushanSTM, KapitulnikSTM}. 

Several years ago, Raghu and co-workers predicted that these surface states should give rise to a new type of collective mode, which they termed a ``spin plasmon"\cite{raghu}. This mode is plasmon-like in the sense that it arises from random-phase approximation (RPA) screening effects, but exhibits a spin current because of the spin-textured character of the surface states. This mode is of both fundamental and practical importance for several reasons. First, its existence is a consequence of electron-electron interactions, and hence is an essential many-body effect in materials that are, traditionally, thought of as independent-electron band insulators\cite{KaneMele, BHZ, zhang2009topological, roy2009, mooreBalents}. Second, recent angle-resolved photoemission (ARPES) studies reported the presence of dispersion kinks in the Dirac quasiparticles in both Bi$_2$Se$_3$ and its superconducting cousin Cu$_x$Bi$_2$Se$_3$, indicating interaction with some bosonic collective mode\cite{KondoKink,wrayKink,chenKink, wray2010spectroscopic,VallaKink}, for which the spin plasmon is a prime candidate. The spin plasmon also has potential application bridging the areas of surface plasmonics and spintronics, by providing a coupling between surface collective modes and spin degrees of freedom\cite{appelbaum,stauber,lai}.

The primary experimental evidence for this mode comes from pioneering infrared transmission measurements of Bi$_2$Se$_3$ nanoribbons, which were fabricated by $e$-beam lithography and reactive ion etching\cite{dipietro,autore,stauber}. An excitation was observed in the THz regime, whose dispersion exhibited the $\sqrt{q}$ dependence expected of the spin plasmon\cite{raghu}. It is crucially important, however, to detect this excitation on a pristine, unpatterned surface, both to corroborate the THz experiment and to understand the dynamics of this excitation in a native material. 

Here, we present measurements of the surface collective modes of Bi$_2$Se$_3$ and Bi$_{0.5}$Sb$_{1.5}$Te$_{3-x}$Se$_x$ (BSTS) using low-energy, momentum-resolved electron energy-loss spectroscopy (M-EELS). M-EELS is an inelastic scattering technique that measures the dynamic structure factor of a surface, $S(\textbf{q},\omega)$\cite{kogar2014temperature}, which is the Fourier transform of the surface density-density correlation function. $S(\textbf{q},\omega)$ is proportional to the bosonic spectral function, $\chi ''(\textbf{q},\omega)$, via the fluctuation-dissipation theorem\cite{kogar2014temperature}. $\chi ''(\textbf{q},\omega)$ directly reveals the charged collective modes of a surface in the meV range, and this makes M-EELS an ideal technique to detect the presence of the spin plasmon in an unpatterned surface. 

To conduct the experiment, single crystals of Bi$_2$Se$_3$ were grown from a melt by techniques described previously\cite{AnalytisSdH}. The Se vapor pressures were varied to adjust the concentration of vacancies, which determine the degree of electron doping\cite{AnalytisSdH}. The crystals were characterized using DC Hall measurements, and labeled 1A-7A in ascending order of bulk carrier density, which ranged from $n_e=1.3\times 10^{18}$ cm$^{-3}$ to $n_e=2\times 10^{19}$ cm$^{-3}$. Single crystals with nominal composition Bi$_{0.5}$Sb$_{1.5}$Te$_{3-x}$Se$_x$ (BSTS) were grown from high-purity (99.9999$\%$) elements of Bi, Sb, Te and Se using the floating-zone method. These crystals are labeled 1B-7B in ascending order of $x$, which was varied over the range $1.5 \le x \le 2.2$. 

\begin{figure}
	\includegraphics[scale=0.37]{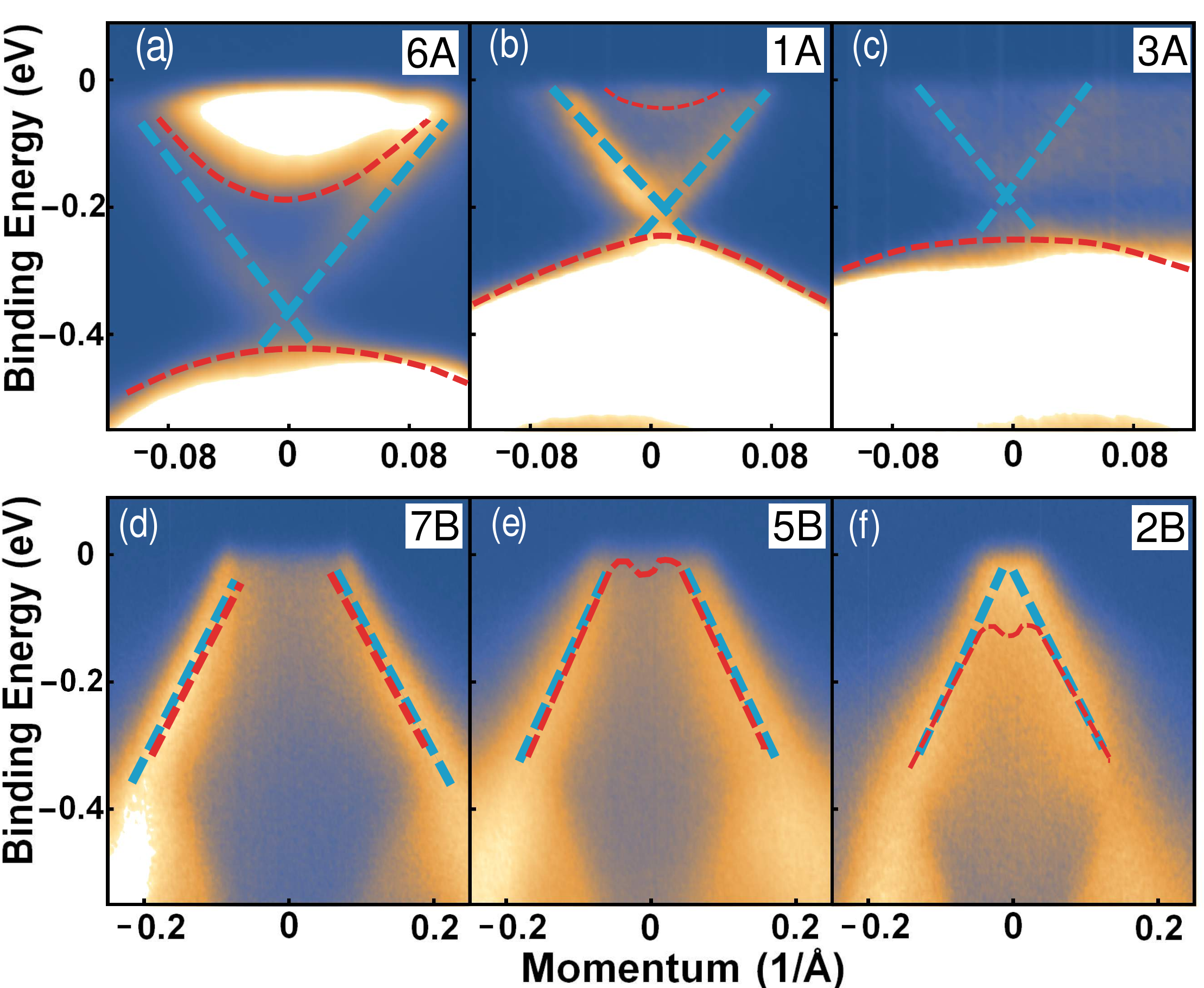}
	\caption{(color online) (a)-(c) ARPES spectra taken at 55K from Bi$_2$Se$_3$ crystals from batches 6A, 1A and 3A, respectively. (d)-(f) APRES spectra taken at 20K of BSTS from batches 7B, 5B and 2B, respectively. The blue dashed lines indicate the dispersions of the surface Dirac bands while the red dashed lines indicate the bulk bands.}
\label{fig:ARPES}
\end{figure}

The crystals were characterized with ARPES to establish the location of the Fermi energy relative to the bulk bands and Dirac surface states. The Fermi energy of some Bi$_2$Se$_3$ crystals in batches 1A-3A was found to reside in the bulk gap, but all crystals from batches 4A-7A were found to have the Fermi energy in the conduction band (Fig. \ref{fig:ARPES}(a)-(c)). We note that, while the bulk carrier density as measured by Hall effect was uniform within a growth batch, the Fermi energy measured with ARPES was highly variable, particularly for crystals with lower vacancy concentration. This variation was observed previously and attributed to band bending due to differences in surface termination upon cleaving\cite{AnalytisSdH}. The Fermi energy in all BSTS crystals was found to reside near the Dirac point, with crystals from batches 3B-7B being slightly $p$-type and those from batches 1B and 2B slightly $n$-type (Fig. \ref{fig:ARPES}(d)-(f)). Note that the valence band in BSTS rises very close to the Dirac point, suggesting bulk carriers may be present even for nearly neutral materials. 

For M-EELS measurements, crystals were cleaved at room temperature in ultrahigh vacuum (UHV) and measured within 30 minutes, unless stated otherwise\cite{supplement, hsieh2009tunable}. The spectrometer used was of the Ibach variety equipped with a double-pass monochromator and an energy analyzer to disperse the scattered electrons onto the detector \cite{ibach1991electron}. To acheive momentum resolution, the spectrometer was equipped with a motorized scattering angle and mated to a custom low-temperature sample goniometer actuated with a piezoelectric motor and differentially pumped rotary feedthrough. Using several sets of translations, the various rotation axes were aligned to intersect the electron beam at a single point. A control system similar to that used in triple axis neutron scattering was employed to allow true reciprocal space scanning. The typical energy resolution was $\sim$10~meV while the momentum resolution was 0.03~\AA$^{-1}$. The incident beam energy was 50~eV for all measurements.

\begin{figure}
	\includegraphics[scale=0.205]{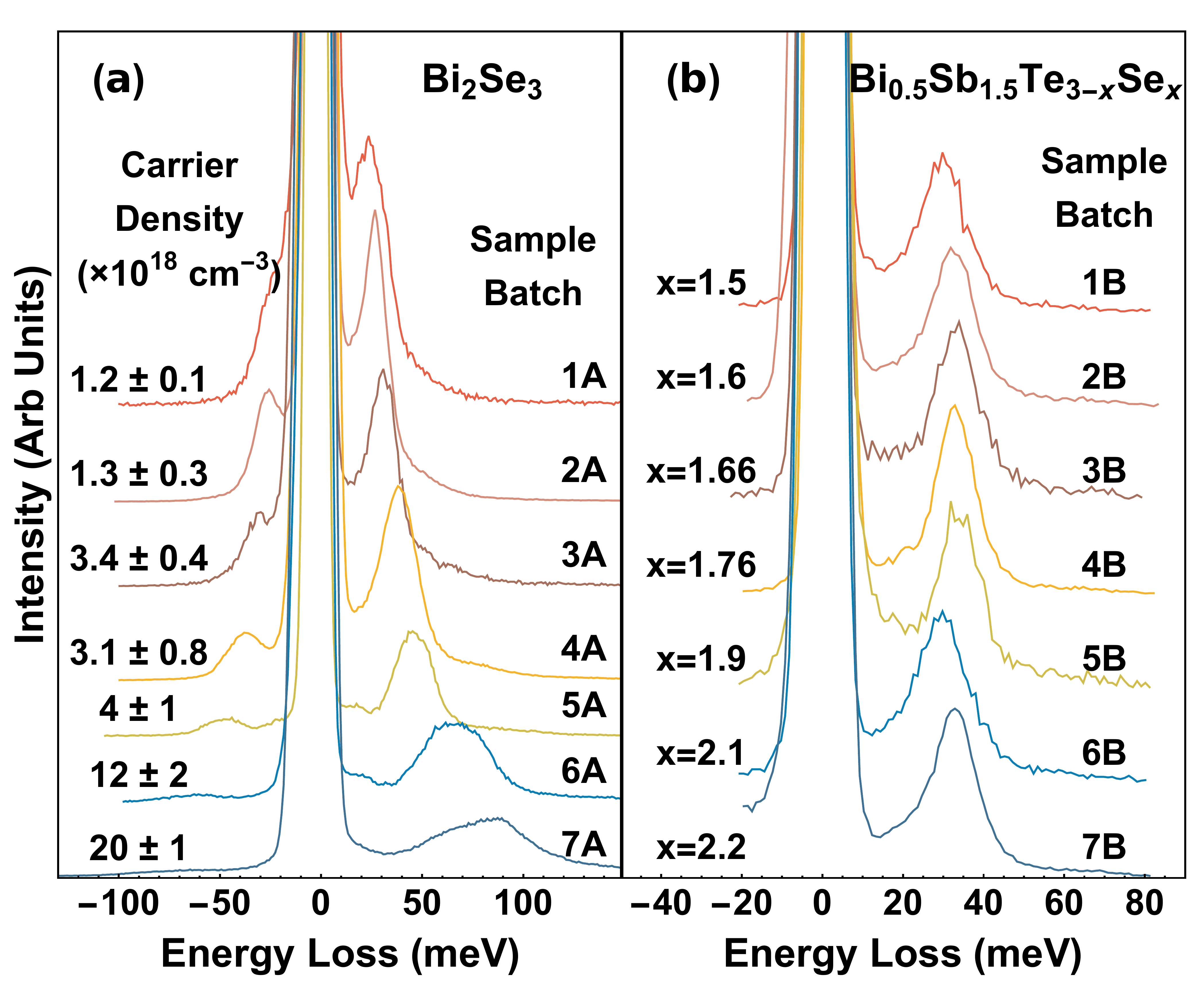}
	\caption{(color online) (a) Room temperature M-EELS spectra from Bi$_2$Se$_3$, taken at \textbf{q}=0, showing the dependence of the plasmon peak on the bulk carrier density. (b) M-EELS spectra from BSTS at \textbf{q}=0 taken at $T=100K$, showing insensitivity of the plasmon peak to the location of the Fermi energy with respect to the Dirac point (spectra are displaced vertically for clarity).}
\label{fig:intensity_doping}
\end{figure}

\begin{figure}
	\includegraphics[scale=0.27]{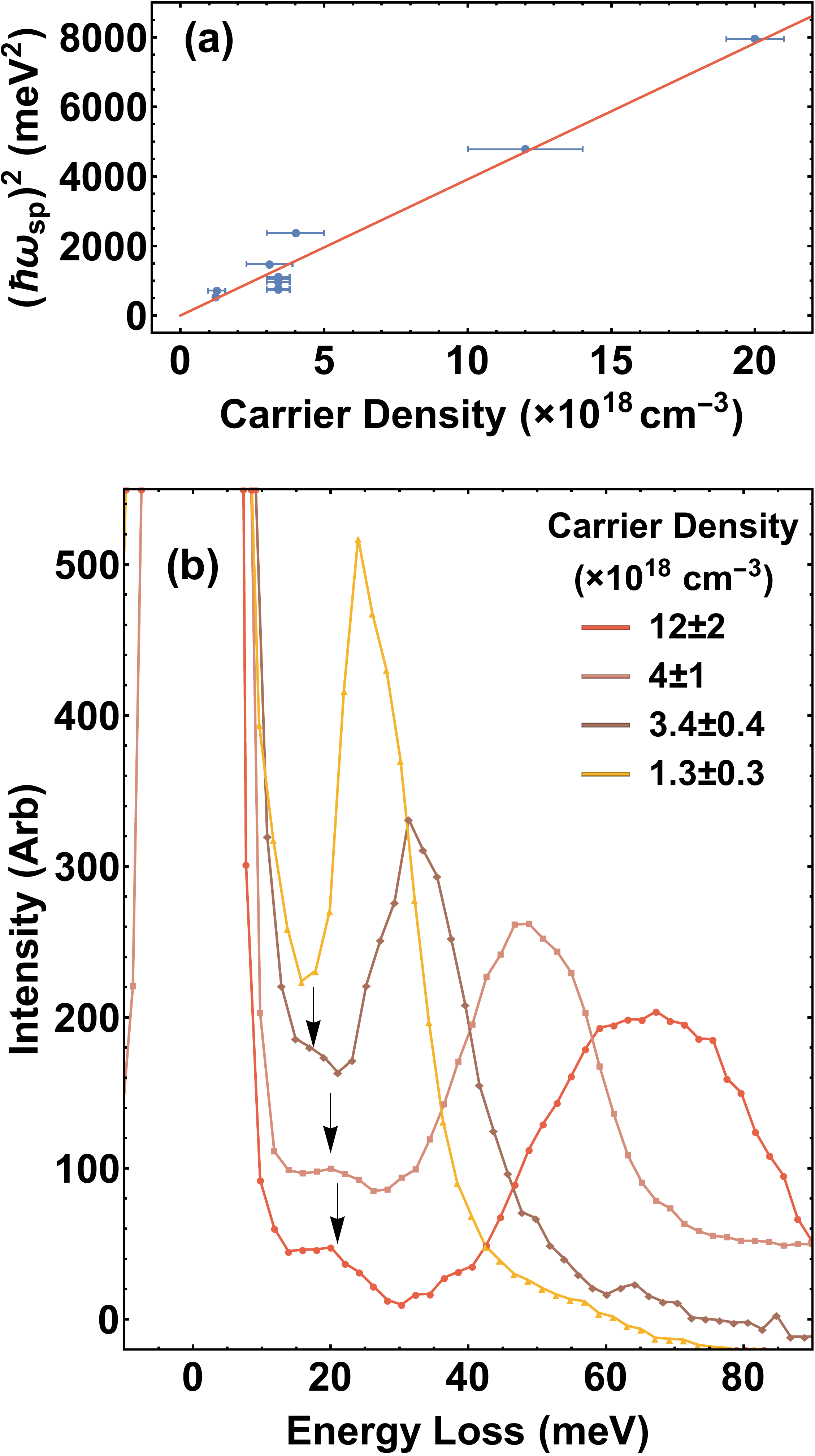}
	\caption{(color online) (a) Square of the surface plasmon energy plotted against the carrier density determined from Hall measurements. (b)Observation of a low energy collective mode, which is the surface analog of the out-of-plane A$_{1g}$ phonon.}
\label{fig:littlePeak}
\end{figure}

M-EELS measurements, taken at room temperature for Bi$_2$Se$_3$ and $T = 100K$ for BSTS, are shown in Fig. \ref{fig:intensity_doping}. The primary feature in both materials is a high-intensity inelastic peak whose energy in the Bi$_2$Se$_3$ system varies from 23 $\sim$ 90 meV, depending upon the bulk carrier density (the peak centered at zero energy is elastic scattering from the crystal surface). In addition, at high doping levels a weak, secondary excitation---with much smaller spectral weight---is observed in Bi$_2$Se$_3$ (Fig. \ref{fig:littlePeak}(b)). This mode may be identified as the out-of-plane A$_{1g}$ phonon previously observed in Raman scattering studies \cite{richter, note1}. 

\begin{figure}
	\includegraphics[scale=0.25]{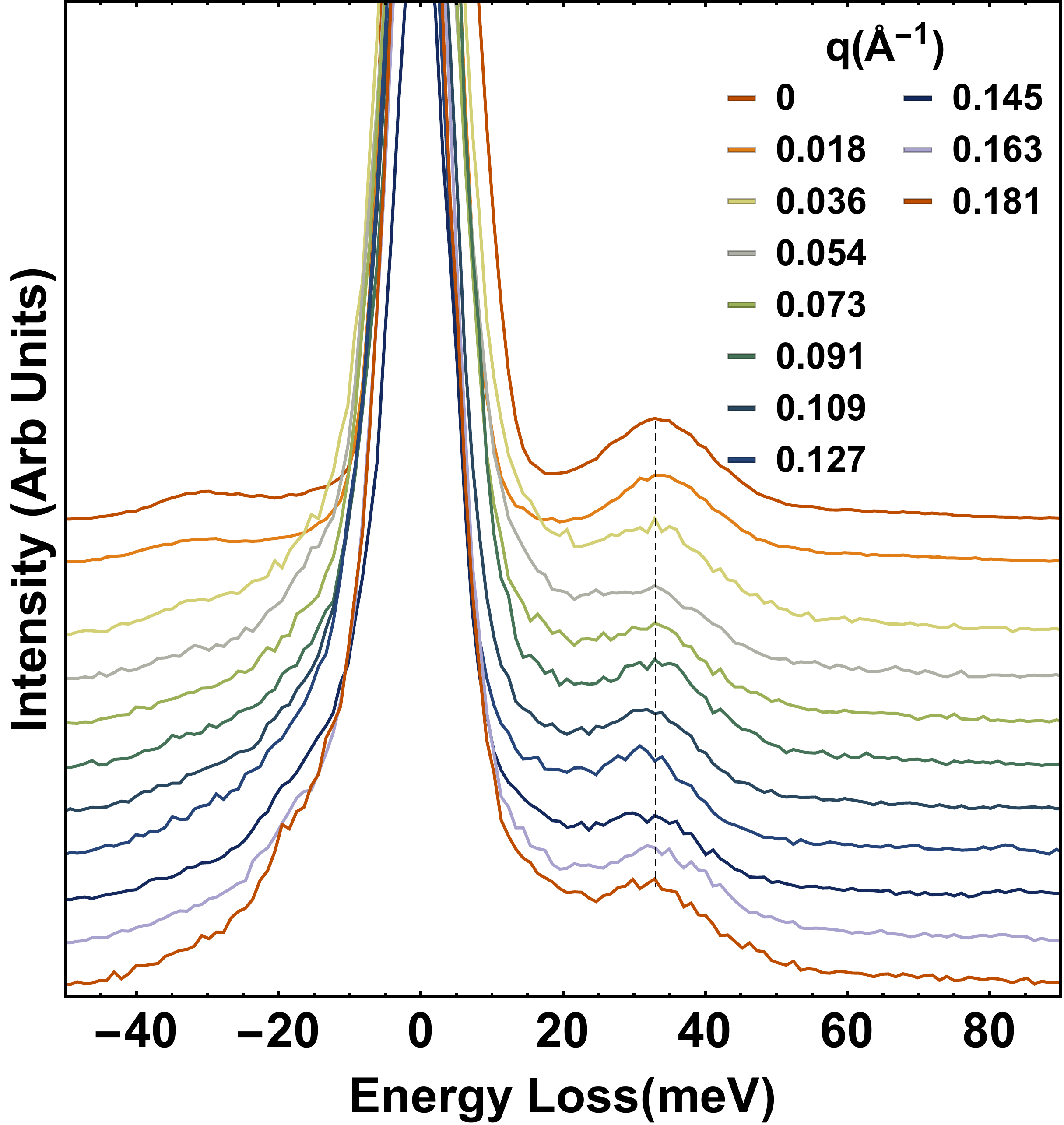}
	\caption{(color online). Momentum-dependent M-EELS spectra from a Bi$_2$Se$_3$ crystal from Batch 3A, showing a lack of dispersion, which is expected of a surface plasmon from a bulk conductor. %Right: Time dependence of a Bi$_2$Se$_3$ sample from Batch 6A compared to a freshly cleaved sample from Batch 7A. Inset: time-dependence of Bi$_2$Se$_3$ sample from Batch 3A.
	}
\label{fig:dispersion}
\end{figure}

While it is tempting to identify the primary excitation as the spin plasmon, its dispersion suggests a different origin. Fig. \ref{fig:dispersion} shows M-EELS spectra for different values of the in-plane momentum transfer, $q$, taken from a Bi$_2$Se$_3$ crystal from batch 3A (other crystals, including BSTS, yielded similar results). Rather than exhibiting the $\sqrt{q}$ dependence expected for a spin plasmon, the peak was found to reside at fixed energy, independent of $q$.

While the peak energy is independent of momentum, it changes in a systematic way with bulk carrier density. Fig. \ref{fig:littlePeak}(a) shows the square of the peak energy plotted against the carrier density, $n_e$, determined from Hall measurements. This plot shows a linear relationship to a high degree of accuracy, indicating that the excitation energy scales like $\sqrt{n_e}$. The reproducibility of this relationship is surprisingly good considering the unpredictable location of the Fermi energy measured with ARPES\cite{AnalytisSdH}, and suggests that the excitation is a feature of the bulk carriers rather than the surface states.

The above behavior is typical of the surface plasmon of a bulk, 3D conductor. In the standard electromagnetic theory\cite{economou,Egdell1992444}, a surface plasmon disperses from zero energy with a phase velocity close to $c$ and, above a momentum, $q_s$, saturates to an energy (in SI units)

\begin{equation}
\omega_{sp}=\left ( \frac{\epsilon_\infty}{\epsilon_\infty+1} \right )^{1/2} \omega_p.
\end{equation}

\noindent
where $\omega_p= \sqrt{n_e e^2 / \epsilon_0\epsilon_\infty m^*}$ is the bulk plasma frequency ($m^*$ being the effective mass), and $\epsilon_\infty$ is a background dielectric constant representing screening by high-energy interband transitions not measured in the experiment. The saturation momentum $q_s \sim \omega_p/\hbar c$ ($\sim 8\times 10^{-6}\AA^{-1}$ for the excitations observed here) is far below the momentum resolution of EELS spectrometers, which therefore observe surface plasmons as nondispersive excitations, as we do here. A least-squares fit of Eq. 1 to the data in Fig. \ref{fig:littlePeak}(a) gives a value $\epsilon_\infty$=26$\pm 2$, which is consistent with previous studies, which usually quote values between 25-29 \cite{richter, dielectric, DrewFarInfrared}. We conclude that this excitation is not the spin plasmon, but a conventional surface plasmon arising from the bulk, conduction electrons in the material. 

Nevertheless, this observation is highly significant. While the spin plasmon should in principle be present, this observation demonstrates that the largest contribution to $\chi''({\bf q},\omega)$ is a surface plasmon of the bulk carriers. The existence of bulk carriers is, of course, well known\cite{AnalytisSdH}. What we have shown is that these carriers can exhibit their own surface collective mode, distinct from any physics related to the Dirac surface states. 

%I can see no reason to do anything other than move this stuff to the supplementary material: 
%To characterize the dependence of the mode on surface quality, time-dependent data were taken for Bi$_2$Se$_3$, summarized in Fig. \ref{fig:dispersion}(b). A spectrum from a sample from batch 6A, which is observed to have a plasmon at 65~meV about a half-hour after cleavage, undergoes a drastic change with time. In a period of 24 hours, the plasmon peak shifts in energy by about 25~meV to 90~meV, yielding a spectrum that resembles a sample from batch 7A, which, nominally, is much more highly electron doped. This data is consistent with observations in ARPES in which the bands were seen to bend at the surface, leading to increased electron doping as a function of time under ultra-high vacuum conditions \cite{hsieh2009tunable}. Time-dependent measurements were also conducted on Bi$_2$Se$_3$ from batch 3A, shown in the inset of Fig. \ref{fig:dispersion}(b). Again, a shift in the peak energy of about 15~meV was observed in the 46-hour period of time examined. This data establishes that the excitation studied here, while deriving from the free carriers of the bulk, is nevertheless highly dependent upon the properties of the surface, which evolve in time in essentially the same manner as observed by ARPES studies \cite{hsieh2009tunable}. Put together, all preceding evidence suggests that the excitation is a conventional surface plasmon arising from the same free carriers that cause bulk conduction.

This surface plasmon is the most likely origin of the quasiparticle dispersion kinks observed in Bi$_2$Se$_3$ with ARPES\cite{KondoKink,wrayKink,chenKink, wray2010spectroscopic,VallaKink}. The coupling strength between a quasiparticle and a bosonic mode is, in the first approximation, determined by the magnitude of the boson propagator, $\chi({\bf q},\omega)$. For a simple electron gas, for example, the quasiparticle self-energy is given by\cite{Ashcroft}

\begin{equation}
\Sigma(\textbf{k},\omega) = - \int_{k<k_F}{\frac{d\textbf{k}'}{(2\pi)^2} V(\textbf{q})(1+V(\textbf{q})\chi'(\textbf{q},\omega))},
\end{equation}

\noindent
where $\textbf{q} = |\textbf{k}-\textbf{k}'|$, $V(\textbf{q})$ is the Coulomb interaction and $\chi'(\textbf{q},\omega)$ is the real part of the boson propagator, which may be obtained from the bosonic spectral function, $\chi''({\bf q},\omega)$, by Kramers-Kronig transform. Hence, those features with the largest spectral weight in M-EELS data are likely to have the largest influence on the quasiparticle self-energy. The dominant feature in our measurements is the surface plasmon from the bulk carriers, whose energy is close to that of the 20~meV ARPES kink. We conclude that it is this surface plasmon, and not the spin plasmon or A$_{1g}$ phonon, that is the origin of this dispersion anomaly.

This conclusion may extend to superconducting Cu$_x$Bi$_2$Se$_3$, which exhibits a dispersion kink at $\sim$90~meV\cite{wrayKink, wray2010spectroscopic}. The Fermi energy from Ref.~\cite{wrayKink} of 250~meV implies a bulk carrier density in Cu$_x$Bi$_2$Se$_3$ of 2.7$\pm$0.8$\times$10$^{19}$~cm$^{-3}$, which (via Eq. 1) implies a surface plasma frequency of 102$\pm$10~meV. This energy is close to that of the observed kink, suggesting that a surface plasmon from bulk carriers is likely the origin of the kinks in superconducting materials as well. 

We close by discussing the implications of our study for optical experiments on Bi$_2$Se$_3$ and related materials. The thickness of the films in the THz study on nanoribbons\cite{dipietro}, for example, were either 60 nm or 120 nm, and hence would act as 2D layers as far as their THz optical properties are concerned, since $\lambda \gg d$. The presence of bulk carriers in these films should therefore give rise to a 2D plasmon that is distinct from the spin plasmon, but also exhibits $\sqrt{q}$ dispersion. A typical bulk carrier density of $n_e=10^{18}$~cm$^{-3}$, for example, would imply an areal density $n_A = 1.2\times 10^{13}$~cm$^{-2}$ for a 120 nm thick film, giving a 2D plasma frequency of $\nu=\sqrt{q e^2 n_A / 8\pi \epsilon_0 \epsilon m^*}$ = 3.5~THz at a momentum of $q=1.6\times10^4$~cm$^{-1}$. This value is close to what is observed in Ref. \cite{dipietro}, indicating that this experiment could, just as well, have been interpreted as observing a plasmon of the bulk carriers. Of course, both the surface plasmon and the spin plasmon should exist, in which case the two would mix in a nontrivial way. Further studies using both M-EELS and THz probes are needed to resolve this issue. 

%[I just don't know where to put this in:] Even at the highest possible doping, $\mu\sim 0.3$~eV, the critical momentum of the plasmon is only $q_c\sim 4\times 10^{-3}\AA^{-1}$. At momenta $q > q_c$ the spin plasmon should be broadened by Landau damping\cite{raghu}. The momentum transfer resolution of most EELS spectrometers $\Delta q \sim 0.01 \AA$, so the spin plasmon would be overdamped on the scale at which the measurements are sensitive. 

In summary, we studied the collective modes on the surface of two topological insulators and found that the primary feature is a surface plasmon arising from the free carriers in the bulk. The A$_{1g}$ phonon is also observed as a secondary excitation with much smaller spectral weight. Because of its large spectral weight contribution to $\chi ''(\textbf{q},\omega)$, this surface plasmon is most likely the origin of the quasiparticle dispersion kinks at 20~meV and 90~meV observed with ARPES in Bi$_2$Se$_3$ and in Cu$_x$Bi$_2$Se$_3$, respectively. This excitation should also exhibit the properties of a 2D plasmon in thin layers, in which it should mix with the spin plasmon, calling for a more nuanced interpretation of recent THz experiments\cite{dipietro,autore,stauber,sim,ou}.

We acknowledge helpful discussions with S. Raghu, A. Karch, S. Gleason, T. Byrum, R. Soto-Garrido, Y. Dai, C. Homes and V. Chua. This work was supported by the Center for Emergent Superconductivity, a DOE Energy Frontier Research Center, under Award Number DE-AC02-98CH10886. Work at Brookhaven was supported by the Office of Basic Energy Sciences, U.S. Department of Energy, grant DE-SC00112704. P.A. acknowledges support from grant GBMF4542 through the EPiQS initiative of the Gordon and Betty Moore Foundation. T.-C.C. acknowledges support from DOE grant DE-FG02-07ER46383. Photoemission work at the Synchrotron Radiation Center was partially supported by NSF grant DMR 13-05583. T.L.H. acknowledges supported from DOE grant DE-SC0012649. G.Y.C. acknowledges support from NSF grant DMR 140871.

\bibliography{bib}

\begin{thebibliography}{36}
\bibitem [{not()}]{note1}%
  \BibitemOpen
  \href@noop {} {}\bibinfo {note} {The smaller restoring force accounts for the
  slight discrepancy in energy of this mode compared to bulk probes, as it
  oscillates out of the plane at the surface.}\BibitemShut {Stop}%
\end{thebibliography}
\end{document}